# Ontological Entities for Planning and Describing Cultural Heritage 3D Models Creation

NICOLA AMICO and ACHILLE FELICETTI, VAST-LAB, PIN, University of Florence, Italy

In the last decades the rapid development of technologies and methodologies in the field of digitization and 3D modelling has led to an increasing proliferation of 3D technologies in the Cultural Heritage domain. Despite the great potential of 3D digital heritage, the "special effects" of 3D may often overwhelm its importance in research. Projects and consortia of scholars have tried to put order in the different fields of application of these technologies, providing guidelines and proposing workflows. The use of computer graphics as an effective methodology for CH research and communication highlighted the need of transparent provenance data to properly document digital assets and understand the degree of scientific quality and reliability of their outcomes. The building and release of provenance knowledge, consisting in the complete formal documentation of each phase of the process, is therefore of fundamental importance to ensure its repeatability and to guarantee the integration and interoperability of the generated metadata on the Semantic Web. This paper proposes a methodology for documenting the planning and creation of 3D models used in archaeology and Cultural Heritage, by means of an application profile based on the CIDOC CRM ecosystem and other international standards.

## 1 INTRODUCTION

3D modelling has become an essential tool for investigation and research in every aspect of the Cultural Heritage field, so much so that its applications have multiplied enormously over the years. From this derives the need to guarantee the quality and trustability of 3D models used as research instruments, largely relying on the timely and detailed documentation of their creation process. To this end is essential the use of adequate procedures capable of guaranteeing the formal description of all the digitisation aspects, from the planning and definition of its general and specific goals and user perspectives to the publishing and distribution of the models, in order to provide details about involved people, times and places, but also about digital instrumentation, environmental conditions, software and parameters present throughout all the stages of the process.

In a previous paper [20] we already investigated this issue and tried to propose a series of guidelines, inspired by the CIDOC CRM [29], for the formal description of 3D modelling performed by means of laser scanners. It was one of the first attempts to define a system for documenting the process of creating a 3D model using the conceptual tools available in 2013. Much progress has been made and many new tools have appeared since then in the field of semantics and ontologies. Some interesting studies were carried out also in the field of photogrammetry, documenting the workflow of digitization [25] and aligning different ontologies with CIDOC-CRM in underwater photogrammetry survey [34].

In this paper we try to extend our field of investigation to any type of cultural digitization and 3D modelling techniques and to define an ontological system, i.e. an application profile, capable of providing a description as complete as possible of the entire process.

Authors' address: Nicola Amico, nicola.amico@pin.unifi.it; Achille Felicetti, achille.felicetti@pin.unifi.it, VAST-LAB, PIN, University of Florence, Piazza Giovanni Ciardi, 25, Prato, PO, Italy, 59013.

Application profiles can be defined as sets of classes and properties used to model the data of a selected domain. Therefore, they are designed for a specific purpose or specific research areas and constructed by selecting, deriving and grouping in a consistent manner the classes and properties of one or more existing ontologies and data schemas, a methodological approach that contrasts with the common practice of building new models from scratch. It is possible to define an application profile by selecting, for example, entities of CIDOC CRM, CRMsci or Dublin Core (or declare new equivalent and related entities) for describing the peculiar objects of the astronomical domain, or the research activities typical of archaeobotany, by means of dedicated models specifically tailored for these purposes. Application profiles of this type have been recently defined for scientific data [43], inscriptions and all other disciplines related to the archaeological research [50] of which ARIADNEplus [4] is aggregating data.

For our model we mainly rely on CIDOC CRM and its extensions since they already provide a large number of conceptual tools to model events, objects, people, places and other elements typical of the 3D modelling field. CIDOC CRM is an ISO standard and represents, to date, the *de-facto* reference ontology for Cultural Heritage.

Compared to what we exposed in our 2013 work, the CIDOC CRM ecosystem, originally designed for a museum environment, has been enriched with many new features and offers now a series of extensions that have been developed over the years for modelling data in many other domains. CRMdig [30, 33], that was partially used in our previous work, still offers enough features for generating provenance metadata about the steps and methods of the digitisation process. For our model we also rely on CRMpe (PARHENOS Entities) [24], a CIDOC CRM extension conceived as a specialisation of CRMdig, developed during the PARTHENOS project for describing the provision, management and use of services, datasets and software by research bodies supplying technological platforms. CRMba [51] is another extension of CIDOC CRM we took into account for the present work, since it is intended to support built archaeology documentation and it can be extremely helpful for relating elements of real buildings and monuments with those represented by 3D.

Our aim in this paper is to lay the foundations and expose the spirit of our new application profile in its general features. More technical details regarding its classes and properties will be provided in more depth in the official technical documentation of the model which will be released shortly. The paper is structured as follows: once the process of creating a 3D model has been outlined (Chapter 2), we provide an overview of our application profile for describing in detail all its phases (Chapter 3) and for defining the mechanisms for the generation, out of it, of the related metadata encoded in formal languages (Chapter 4) before moving to conclusions (Chapter 5). The Appendix at the end of the paper provides a brief overview of the newly introduced classes and properties.

## 2 DEFINING THE PROCESS

### 2.1 Overview on digitization process

The complexity and heterogeneity of digitization and 3D modelling in Cultural Heritage produced a wide range workflow and guidelines [2, 13, 19, 35, 49]. To ensure the scholarly quality of the models and to avoid the loss of knowledge, it is crucial to find a common framework, generally accepted, related to with quality assurance.

Numerous projects and documents aimed at creating guidelines and good practices. European projects such as 3D COFORM [28], 3D ICONS [16] and CARARE [5], to name a few, have contributed in the development of tools and definition of guidelines for supporting the digitization and visualization methodologies of 3D asset. In the field of computer-based visualization the London Charter [32] an internationally recognized set of principles ensures the intellectual and technical rigor in the reconstruction and visualization of 3D dataset in Cultural Heritage domain. The London Charter is the theoretical framework of The Seville Principles [42] that marks the best practices in computer based archaeological visualisation. Despite these efforts in the direction of defining principles and guidelines in 3D reconstruction and visualization, we are still far from their full application in the Cultural Heritage field. In particular, the development of a common accepted workflow is lacking. Although there is no universal workflow for capturing cultural resources in 3D, there are similarities among the different methods and workflows. Following the study of Pfarr-Harfst [47], that identified four broad phases in the digitization pipeline, i.e. preparation, data collection, data processing and publication, we have updated, extended and defined a new approach for the management of the digitization and 3D modelling process in CH field (Figure1).

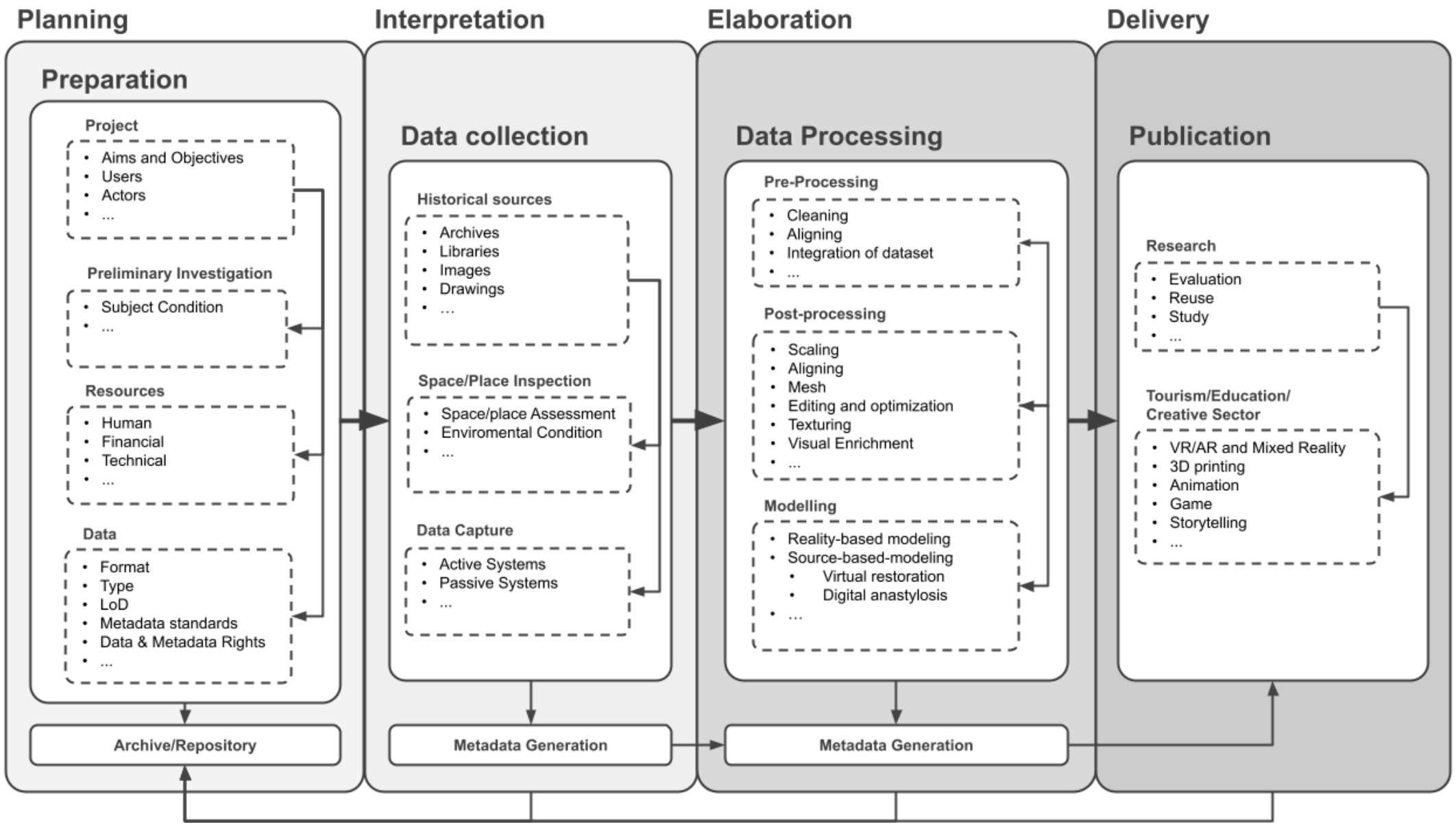

Fig. 1. Digitization and 3D modelling workflow.

## 2.2 Preparation

The preparation phase is crucial to define aims and objectives of digitization, to address all the human and financial resources, technical infrastructure etc., necessary to achieve the desired results. The definition of the goals also includes an agreement with the other parties involved in the project, regarding the rights of the documentation produced [23].

This section determines moreover the creation of a repository, with the definition of the file naming, file formats, versioning, metadata and paradata, as well as storage and backup strategies. This information is usually contained in a dedicated document, the Data Management Plan, which also includes information on how data will be made accessible, how data will be licensed to allow re-use, if there are restriction of use and so forth.

The creation of digital archives is a pivotal step to collect all the documentation produced by the project and to provide access to digital records of data, documentation, interpretation processes and analyses. A recent study on commercial and institutional repositories states that a number of repositories fails to provide related information and resource links for further study and reuse [27, 40, 52]. Most of the 3D objects repositories use an external viewer and rarely online editors to perform quantitative and qualitative analysis on 3D data or does not offer diverse version packages of the same file, usable for different purposes.

Within the repository, all the information gathered during the different phases of the workflow should be collected following the FAIR principles, ensuring access, readability and reliability of the dataset.

In some cases, where the digitization process does not generate standard file formats, it would need to develop trustworthy procedures for normalizing or converting file formats compliant with accepted standards, and possibly store the operating system and the software used, to ensure data long-term preservation [40].

### 2.3 Data collection

Beyond the creation of a 3D model a large amount of documentation is collected, coming from different research fields: architecture, archaeology, history of art, conservation etc. The data collecting phase includes the quantitative and qualitative data retrieval.

Textual and iconographic sources such as written descriptions, conservator's reports, architectonic and artistic paintings, drawings and sketches, images and comparison with similar contexts must be analysed and interpreted before starting the data capture and 3D modelling.

The data capture and the selection of devices and techniques start on the basis of information obtained from the documentation and from the inspection of the space/place where the digitization will be performed.

Subsequently the interpretation and analysis of all the data gathered from the collected documentation introduces the data processing.

### 2.4 Data processing

In our previous work we analysed the digitization process and 3D data processing based on data capture of the reality, focusing on laser scanning technology. To find a common framework and consider all possible applications, here it was adopted a broader approach to the argument, which does not pretend to be exhaustive.

Data processing extended to digitization and 3D modelling implies a differentiation between a reality-based modelling [39], based on the digital acquisition through 3D survey of existing contexts and source-based modelling [21], virtual reconstruction of highly damaged or lost contexts.

Within the reality-based modelling after the digitization of the object and before starting the 3D modelling, the first step of data processing includes the data registration, that can be automatic or semiautomatic. It refers also to the integration between different datasets, created during the data collection (e.g., photogrammetric and laser scanner data). For each dataset, created with diverse devices or techniques, different processing procedures are performed (e.g., cleaning the range maps from noisy data) in order to avoid redundant and noisy outputs.

The results of the registration process are used as point cloud to generate different outcomes or processed with different software. After registration, the point cloud is converted into a polygonal model (mesh). The accuracy of the final outcome is affected by the parameters defined by the operator during the processing phase. For on-line visualization, it may be necessary to apply dataset decimation and resampling to reduce the file size and the geometric complexity of the 3D model. Visual enrichment is the final step of the workflow, used to obtain a photorealistic 3D model.

Source-based modelling relies on the analysis and the interpretation of the documentation and historical records available in archives and libraries. These two approaches could be carried out simultaneously, producing multiple 3D models from source and reality, to develop an interpretative study and a comparative analysis of the data collected from the reality and from historical and iconographic documents [21].

### 2.5 Publication

Once 3D models have been produced, documented, and archived locally, there is the need to find the best way to reach identified audiences: scholars, student, the public at large etc. There is a large amount of possible application of 3D digital output. The widespread use is interactive 3D online browsing and visualization, followed by AR/VR and Mixed Reality using different devices, 3D printing and all the derivative content such as animation, serious games, movies created by the creative sector involved in the exploitation of cultural content. Visualization of 3D asset should follow the guidelines of the London Charter and the Seville Principles to avoid misinterpretation of the visualized results as historical truth, due for example to the photorealism of the 3D model.

A strategy to disseminate 3D content, defined in the preparation phase should identify more than one platform or existing online aggregator infrastructure [3, 6, 8] that delivers the same 3D content in as many ways as possible.

In recent years different online platforms were developed, with different functionality related to the technology used and to the outcomes: point cloud, textured model, animated models and file for 3D printing [1, 7, 10, 12, 14, 55]. Choosing the right platform or developing a new one, with different tools, depends on the project aims and must be identified during the preparation phase.

For preservation and usability purposes, it is also necessary to provide suitable storage and publication services that guarantee, in compliance with the FAIR principles [56], the findability and accessibility of 3D models, but also the possibility to view and use them in controlled collaborative environments or virtual labs through appropriate dedicated services that allow to carry out most of the operations described in this paper, and among others:

- Different ways of visualizing the 3D models available in the archive.
- The annotation and enrichment of the models or parts of them.
- The reusability and the derivation of new models starting from the available ones.
- The elaboration and generation of new models, including composite ones, built by combining existing ones.
- The publication of the developed models and the related information.
- The sharing of all data and metadata with other users of the system.

Platforms of this type, collaborative and equipped with a wide range of dedicated services, certainly require a complex infrastructure that is also under investigation by our team and is planned to be built in a close future. The Cloud represents a privileged tool in this sense since it provides articulated and complete solutions for data and metadata management and for their efficient use.

Once the suitable platforms for the project have been identified, the 3D model must be exported in the right format avoiding loss of data (raw data must be archived and made available upon request of researchers and professionals).

In computer graphics a large number of file formats, both proprietary and open, are produced. The digitization workflow generates several dataset, that can be processed connected to the final publication: online visualization, AR/VR and Mixed reality, gaming and 3D printing. For each of these outcomes it is necessary to export in the dedicated format. For online visualization and archival .X3D, .OBJ .PLY are commonly used in the heritage community. Among these, the CNR-ISTI Lab developed the multi-resolution .NEXUS file format [11] used for the 3DHOP viewer, designed for visualization of reality-based modelling output only. WebXR (Web eXtended Reality) and glTF are file formats designed for delivering VR and AR to devices, including head-mounted displays. Recently the open interchange format glTF has been rising in prominence because it could be used to transfer the whole 3D scene and support animation, material and color mapping.

Beside the processing and exporting of modelled or digitized data, it is necessary to define also the standard format for displaying the metadata and paradata.

These platforms offer some tools for annotating, enhancing the quality of rendering, creating storytellings and animations by selecting some parameters offered by the platform. In addition, platforms such as Sketchfab offer the possibility to engage discussion with the users. Other platforms are specialized in 3D printing and give the possibility to upload and process 3D printable representations of cultural artefacts [10]. An evaluation of scientific quality of 3D on-line platforms was carried out by Statham [52], but there is still some concern regarding the management of large-scale data, tools for analysis and evaluation, render fidelity [9], level of documentation and engagement.

# 3 MODELLING THE PROCESS

## 3.1 Overview of the modelling principles adopted

The pipeline described in the previous chapter can be modelled employing the classes and properties we have chosen to build our application profile. This section presents all the entities mentioned and the semantic formulation of the different steps in which they are involved. For the construction of our application profile we defined new specific classes when necessary, and used the classes provided by existing ontologies when the modelling requirements of the investigated entities were already met by one or more of them. Table 1, below, shows all the ontological models used for the definition of our profile, with information about their version and the prefixes used to identify their classes and properties.

Table 1. Ontological models used for the 3D application profile, with version, classes and properties prefixes and description.

| Model | Version | Prefixes | Description |
|---|---|---|---|
| CIDOC CRM | 7.1 | E / P | A formal ontology for Cultural Heritage information |
| CRMsci | 1.2.9 | S / O | The scientific observation model |
| CRMdig | 3.2.1 | D / L | Model for provenance metadata |
| CRMpe | 3.1 | PE / PP | The PARTHENOS Entities model |
| CRMba | 1.4 | B /BP | Extension to support buildings archaeology documentation |

An overview of the new classes and properties introduced by our application profile is provided in Table 2 and Table 3. The prefixes "3DE" and "3DP" were chosen to identify them. Details about their nature and derivation are provided in Appendix, while indications about their use in relation to the specific scenarios outlined for this paper are provided in the following sections. The super classes and super properties of each of them are reported in italics below their names in the graphical representations presented in this chapter.

Table 2. New classes of the 3D application profile.

| Class | Subclass of | Represents |
|---|---|---|
| 3DE1 Digital Model Creation | E7 Activity (CIDOC CRM) | Activities of creation of a 3D model |
| 3DE2 Digitisation Campaign | E7 Activity (CIDOC CRM) | Campaigns of 3D digitisation |
| 3DE3 Digitisation Session | E7 Activity (CIDOC CRM) | Sessions of 3D digitisation |
| 3DE4 3D Model | D1 Digital Object (CRMdig) | The 3D models in their digital nature |

Table 3. New properties of the 3D application profile.

| Property | Subproperty of | Relates |
|---|---|---|
| 3DP1 was made during | P10 falls within (CIDOC CRM) | 3DE1 –> 3DE3 |
| 3DP2 has session | P9 consists of (CIDOC CRM) | 3DE2 –> 3DE3 |
| 3DP3 is managed by | PP4 is hosted by (CRMpe) | 3DE4 –> PE2 |

The sections of this chapter propose a conceptual rendering of the corresponding sections analysed in chapter 2, each focusing on a specific stage of the process. The diagram shown in Figure 2 serves to illustrate the distribution of the main entities involved in the 3D modelling process along the various phases of the workflow previously described. The representations proposed in the other figures of this chapter, on the other hand, follow an ontological approach and serve to show how, regardless of the workflow, the same entities are linked through the hierarchical and semantic relationships provided by the model in order to give an idea of how the conceptual Knowledge Graph is created.

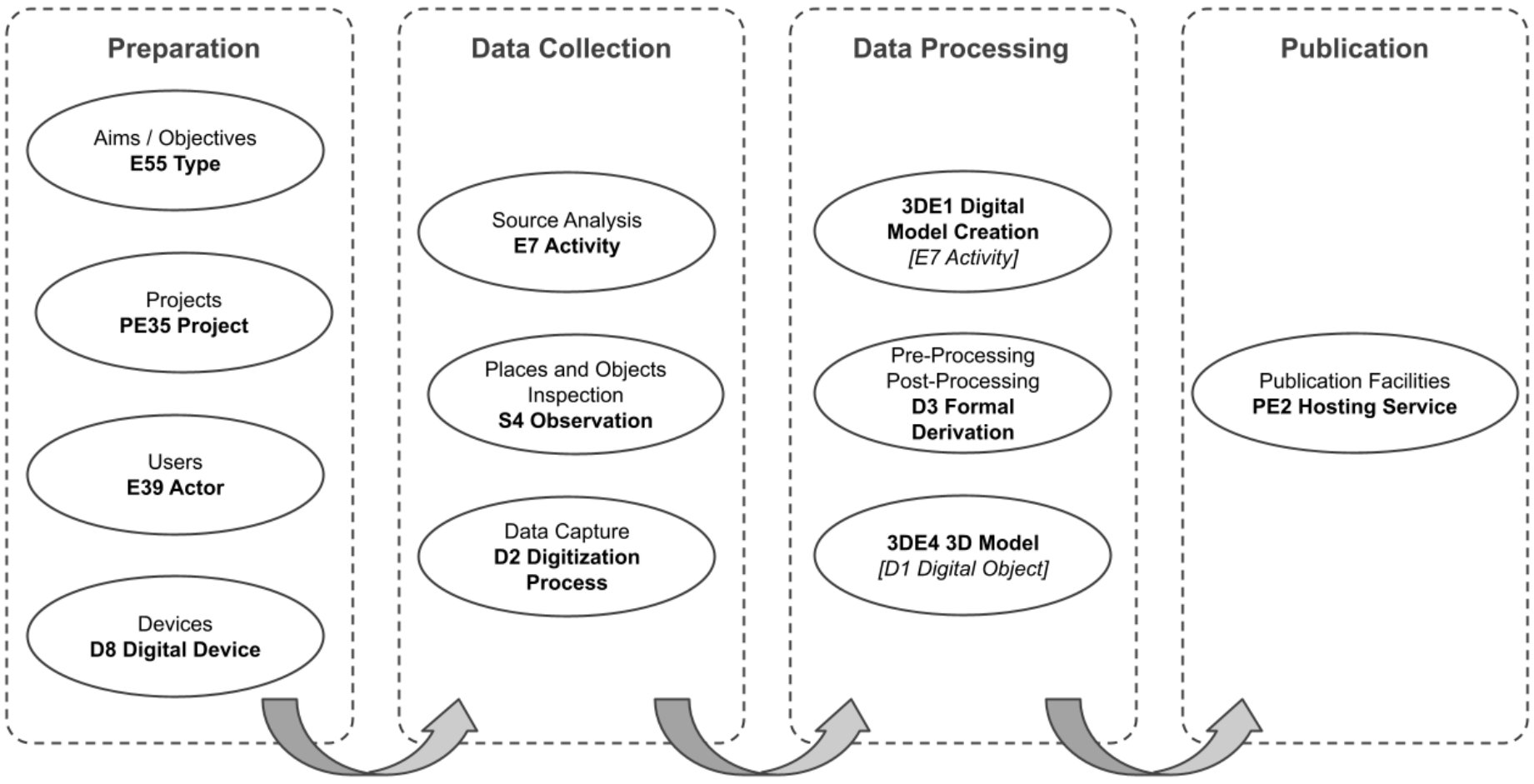

Fig. 2. Overview of the main classes involved in the various steps of the workflow.

## 3.2 Modelling the preparation phase

*3.2.1 General and specific goals.* As we already underlined, the definition of the purposes and scientific questions for which 3D models are created is a fundamental part of their documentation since all subsequent activities depend on it. Very often, the specific purposes of the acquisition are linked to the general goals of the project within which they are carried out.

To describe the set of actions necessary for the creation of a 3D model we have introduced the new *3DE1 Digital Model Creation* class, which specializes the *E7 Activity* class of the CIDOC CRM to be more adherent to the specific actions required by the digitization process. The *PP43 supports activity* property of CRMpe is ideal for relating these digitization activities to the specific purposes of the research project (*PE35*) within which they take place.

The *P20 has specific purpose* and *P21 had general purpose* properties can be used to describe the mutual relationships between all these activities, in order to define, for instance, the main goals of a project or the reasons why a certain 3D model is created (for which, the *E55 Type* class is employed). The *P70 is documented in* property can be used to link all the documentation produced during the various phases of the digitisation process.

All these entities and the way they are linked are shown in Figure 3.

*3.2.2 Institutions, people and types.* Users, institutions and other actors involved in various ways in the different activities, can be modelled by using instances of the *E39 Actor* class and its subclasses (*E21 Person*, *E74 Group*, *PE34 Team*, *PE25 RI Consortium*) along the whole process. CIDOC CRM already provides all the necessary properties to relate actors and activities in many ways. The *P14 carried out by* and the *P11 had participant* properties are used to distinguish who actually performed the digitisation work from other indirect or incidental participants. The same applies to typologies, modelled in CIDOC CRM through the *E55 Type* class, that can be assigned to any entity of the model through the *P2 has type* property. The concepts for defining typologies can be standardized through controlled vocabularies defined locally or shared on the Web.

*3.2.3 Projects, campaigns and sessions.* Very often, the digitization activities planned and promoted in the framework of research projects are organized in campaigns and sessions of acquisition and processing activities that need to be documented in order to provide completeness to metadata. We can define a project as a collaborative initiative undertaken over a period of time by groups of actors with the intention of performing a defined 3D digitisation program. Campaigns and sessions are sets of planned activities taking place within a defined space and time and performed by groups of actors who cooperate in a coordinated way. There is obviously a close relationship between these elements as one or more sessions can take place within a campaign and both campaigns and sessions can be part of the activities of a project.

From an ontological perspective, a project can be modelled using the *PE35 Project* class of CRMpe. This class is intended to describe "collaborative enterprise undertaken over a period of time by an instance of *PE34 Team* with the intention of effectuating some defined program entailing the support of a number of instances of *E7 Activity*", as stated in CRMpe documentation. In this sense, a *PE35 Project* may be a part of what is usually called "project", which may consist of various distinct sub-projects, sometimes called work packages, each one being a *PE35 Project* and formed by various coordinated activities. For example, the project of documenting a building may consist of several sub-projects documenting its parts. Thus, the related *PE34 Team* can be also used to model the group of researchers working together in a project and carrying out its various activities. For modelling campaigns and sessions we have introduced the new classes *3DE2 Digitisation Campaign* and *3DE3 Digitisation Session*, subclasses of *E7 Activity*, and we have put all these entities in a reciprocal relationship through the *P9 consists of* and the newly defined *3DP2 has session* properties. The role of these actors in the preparation activities for digitization operations and the different entities involved are shown in Figure 3.

## 3.3 Modelling data collection

Given the complexity of the work of preparing the documentation and the conditions within which the modelling will take place, it is of fundamental importance to structure all the activities required by this phase in a coherent way. The newly introduced “Data Collection” phase extends the "Location Survey" phase described in our previous work [20] to comprise, in addition to the inspections and the assessment of the places and the environmental conditions, the collection, analysis and interpretation of the existing historical documentation available in archives and bibliographic sources, necessary for planning effective works. It also comprises the assessment of the status and physical conditions of the objects to be scanned. From an ontological point of view, it encompasses a coordinated set of activities aimed at acquiring material, geographical and historical knowledge of heritage buildings, monuments and objects of which the 3D modelling will be carried out and produce reports and other documentation to be used during the subsequent steps of the process.

*3.3.1 Inspections and observations.* To model all the activities forming part of the general one documented by the *3DE1 Digital Model Creation* class, it is possible to define instances of the *E7 Activity* class and use the related properties *P9 Consists of* to describe how all of them take place. We also introduce the use of the *S4 Observation* class of the CRMsci extension to model the specific assessment activities of the physical characteristics of the objects and the environmental conditions of the places where they are located (e.g., the site of monuments, buildings, archaeological excavations) and of the buildings and the other places where they are kept (e.g., laboratories, museums etc.).

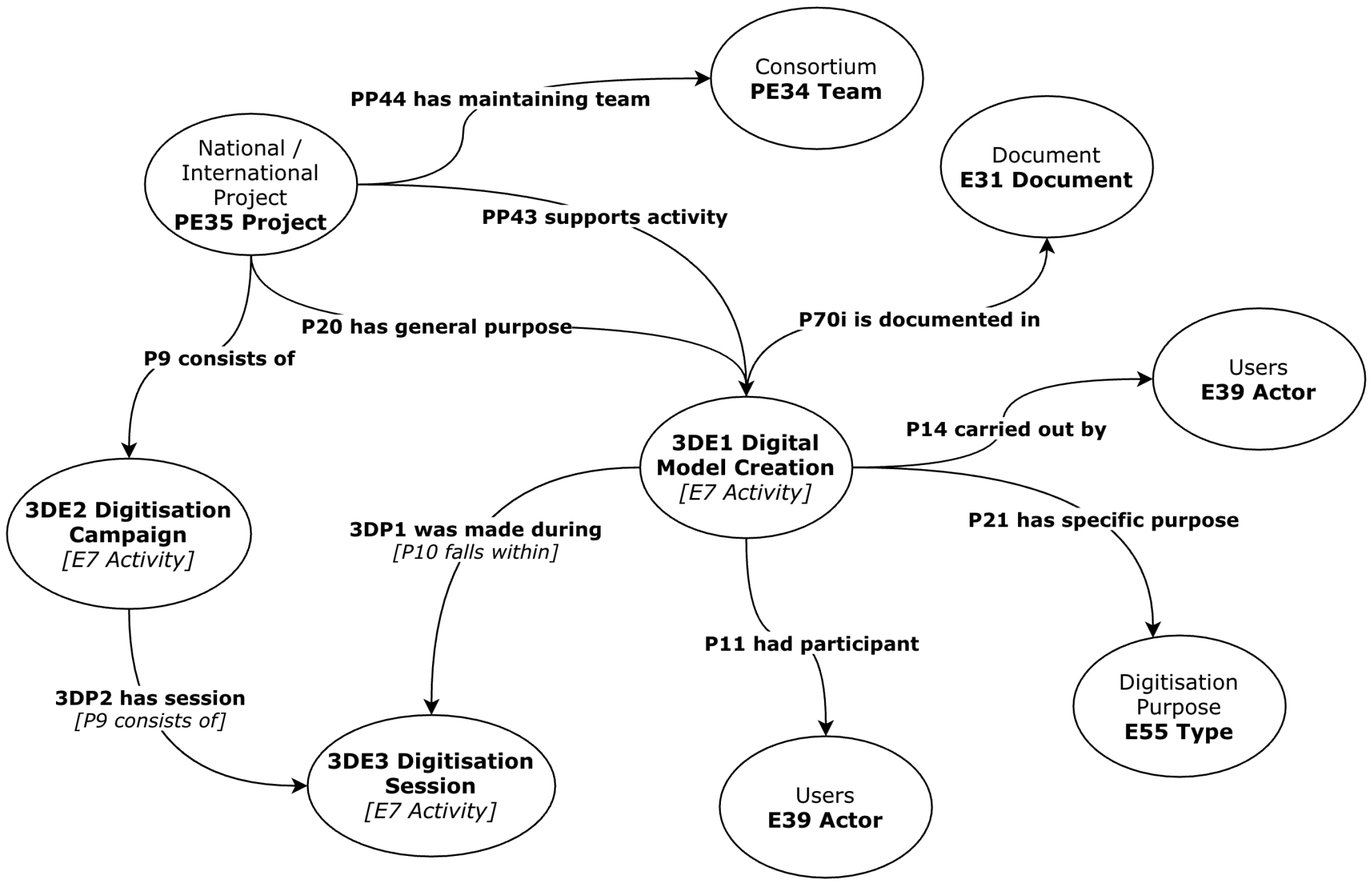

Fig. 3. General overview of the entities and properties used for describing the preparation phases and the involved actors.

If necessary, instances of *S4* can also be defined for the specific investigation of documents such as maps, drawings, reliefs and similar, used for the virtual reconstruction of disappeared or transformed environments. The properties *O8 observed*, *O9 observed property type* and *O16 observed value*, having *S4* as a domain, can be used to specify, for each of these entities, how each investigation took place.

*3.3.2 Creating documentation.* The results of location surveys and source analysis are usually recorded in various documents, reports and other documentation (*E31*) whose preparation is modelled by means of the class *E65 Creation*. The tight relationship between these activities and the created documentation is expressed through the *P20 had specific purpose* property.

The whole process of preliminary inspections, information preparation and data collection is illustrated in Figure 4.

*3.3.3 Digital acquisition and model development.* Digital acquisition activities are rendered through the *D2 Digitisation Process* class of CRMdig and its related properties. In the case of a reality-based model, the *L1 digitised* is used for specifying the scanned object, monument or site (*E18*), and the *L12 happened on device* to describe the device used during the scanning phases (*D8*).

Other devices and add-ons used together with digital tools for the acquisition activity are documented using the *E18 Physical Object* class and are linked to the digitization process by the *P16 used specific object* property and to the digital device used (*D8*) using the *P46 forms part of* property. The *L13 used parameters* property can be used to detail the specific settings employed by the process.

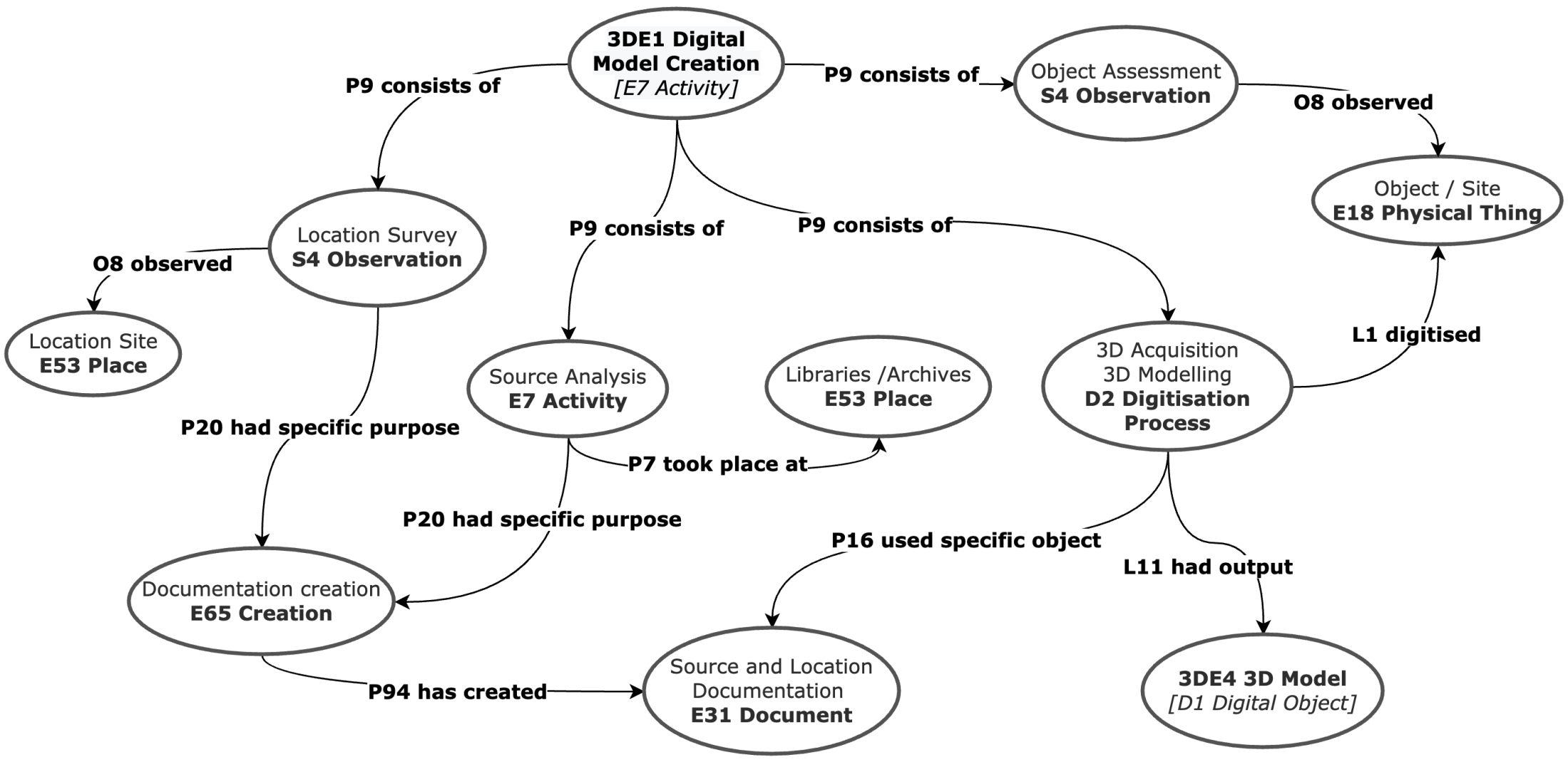

Fig. 4. General overview of the entities and properties used for describing the preparation data collection phase.

In the event that the model has been generated in a procedural way, i.e., starting from the information obtained from site surveys and archive research, the *P16 used specific object* property can be deployed to relate the 3D development activity to the documentation used as the basis for its performance.

In any case, the *D14 Software* class and the related *L23 used software* property can be used to specify the applications, software codes, computer programs, procedures and functions that are employed to generate and operate a system of 3D objects. Finally, the *L30 has operator* property can be used for identifying the technical operators who actually performed the digitisation activity. The new *3DE4 3D Model* class, subclass of *D1 Digital Object*, has been introduced to represent the 3D models resulting from the scanning and the modelling operations. The *L11 had output* property is used to link a digital object to the event that created it.

A general overview of the activities composing the digital acquisition phase is presented in Figure 5.

## 3.4 Modelling data processing

Data registration, post-processing and other elaboration performed on 3D models can be understood, from an ontological point of view, as a series of actions that, using as input a certain digital object, generate a new object preserving the representation features of the original one but in a different form, as a result of the various processing, transformations and integration performed. Being the new object a *3DE4 3D Model* as well, and thus being linked to the original one by a "derivation" relationship, the class more suitable to be used for this activity is the *D3 Formal Derivation* of CRMdig.

The two related properties *L21 used as derivation source* and *L22 created derivative* are also ideal to detail the process of transition from the original digital object to the processed one. The same conceptual model can be deployed multiple times also in order to describe the procedure of integrating two or more dataset into a new one. The *E55 Type* class and the relative *P2 has type* property can be conveniently employed in all these cases to specify the format of the resulting 3D model (e.g., "Point Cloud", "Polygonal Mesh" etc.). A similar mechanism can be used in combination with the *D3 Formal Derivation* class to specify the type of intervention performed on the 3D model (e.g., "Decimation", "Visual Enrichment" and so on) to obtain a certain result.

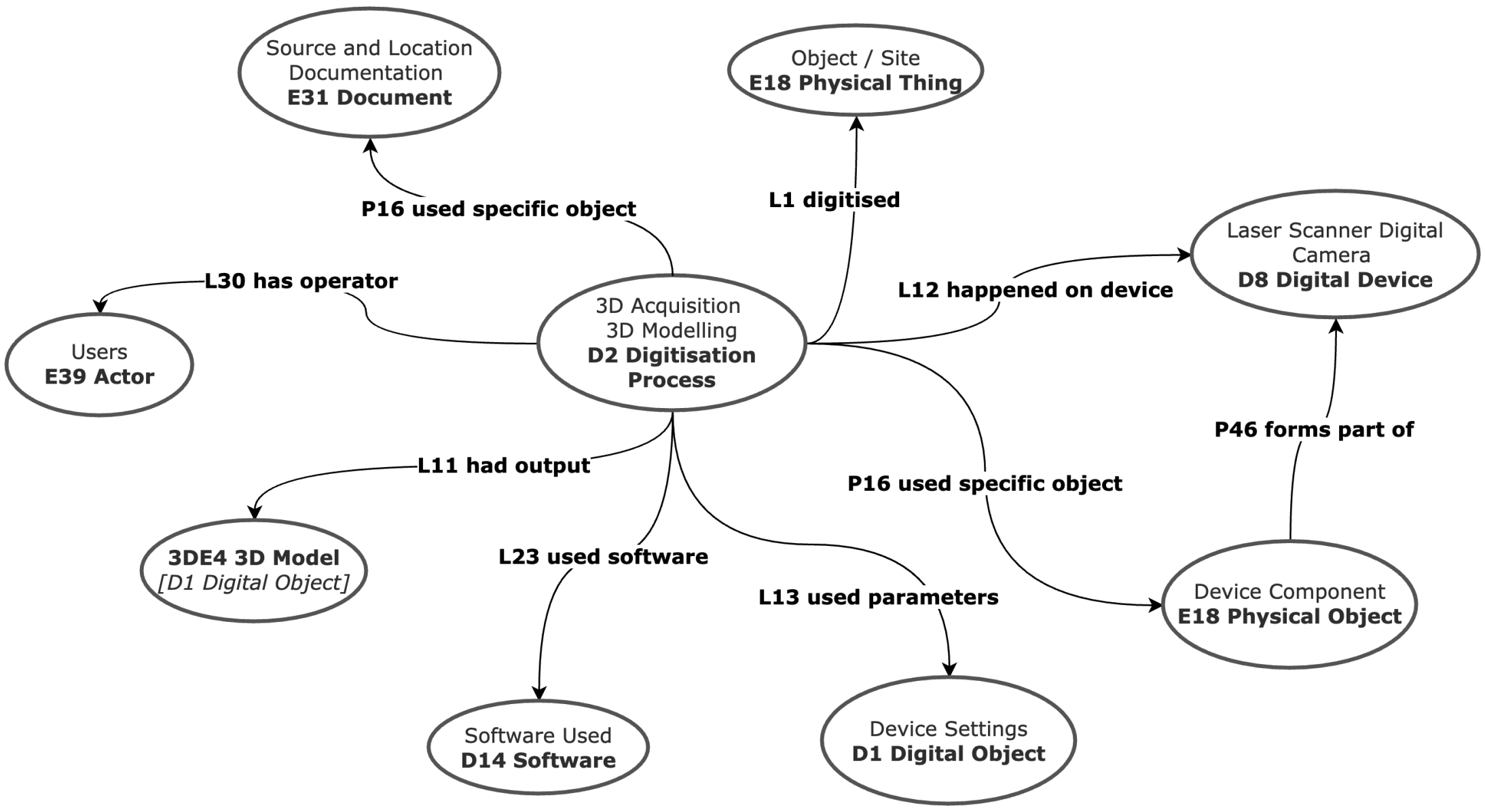

Fig. 5. Overview of the 3D acquisition and modelling activities.

The specific software used can be also documented employing the *D14* class, as in the case of the acquisition and modelling steps. The *P33 used specific technique* property is ideal to describe the various techniques used for the elaborations, while the *L13 used parameters* property offers the possibility to specify the parameters chosen by the operator to perform the various operations. The data processing modelling is shown in Figure 6.

## 3.5 Modelling publication

CRMpe confers to our model the ability to fully describe the process of publishing 3D data and other related information on generic or dedicated repositories via the *PE2 Hosting Service* class, in combination with the new *3DP3 is managed by* sub property of the *PP4 is hosted by* property. This new property allows modelling not only the concept of storing a 3D model on a repository, but also the ability to browse and perform other operations on it while it is hosted on dedicated platforms. More information about the repository can be added using the *PP2 provided by* property, giving indications about the actors (*E39*) providing the hosting service, and the *PP49 provided access point*, indicating the URL address, APIs access or the location of other digital places (*PE29 Access Point*) where the repository can be reached and its services used. The *PP50 accessible at* property can, in turn, be deployed to state a direct link between the access point and the 3D models available through it.

The use of software already available on the same repository, as specified by means of the *PP4* property, can be modelled by linking instances of the classes and the properties provided by CRMdig, specifically the *D10 Software Execution*, that together with *L23 used software* and *L2 used as source* are particularly suitable for describing the 3D models visualisation, annotation, processing and transformation capabilities that a repository may offer as additional features. Figure 7 provides a general overview of how the publication mechanism is rendered by our model.

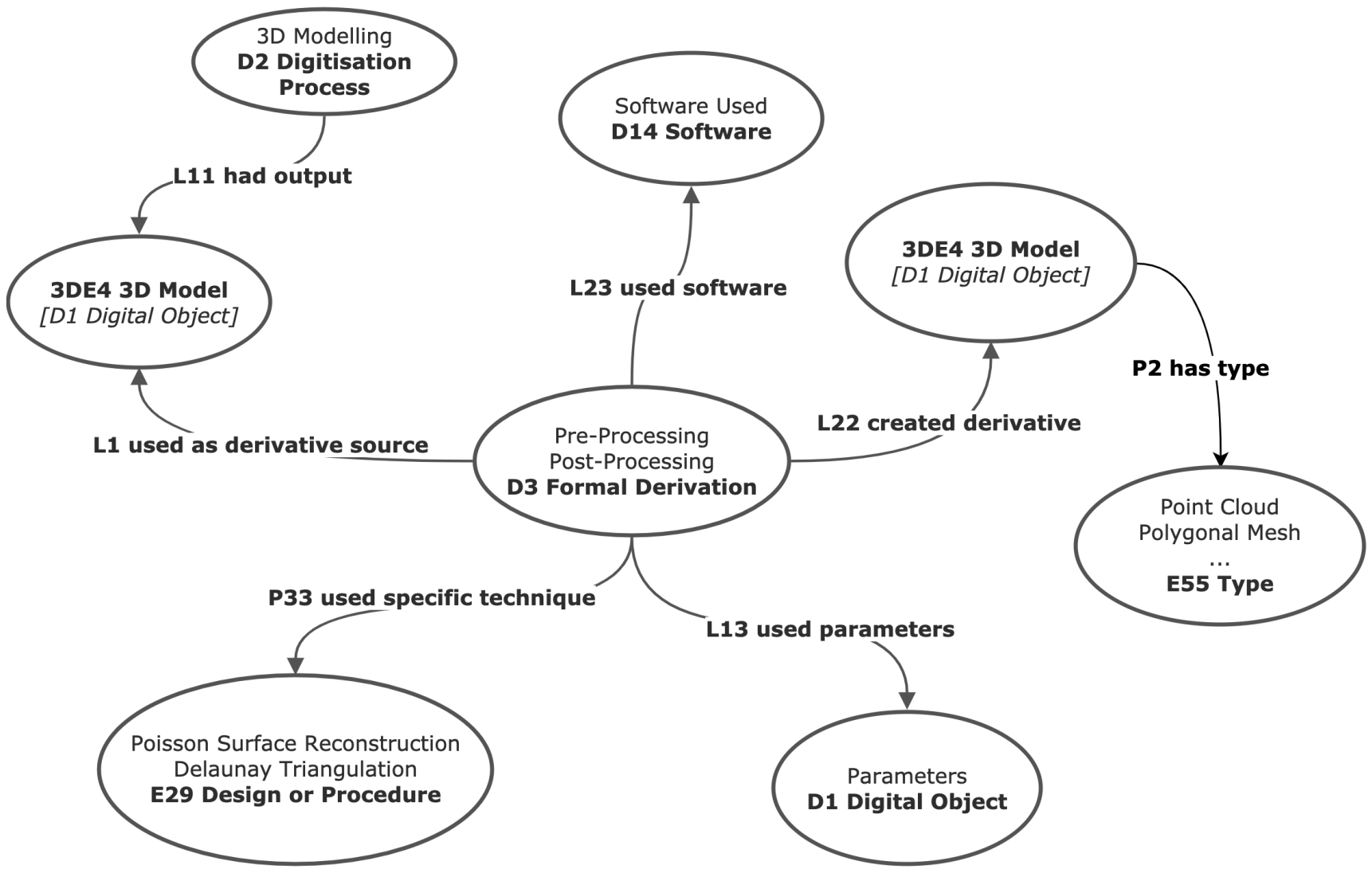

Fig. 6. General overview of the entities and properties used for describing the 3D data processing phase.

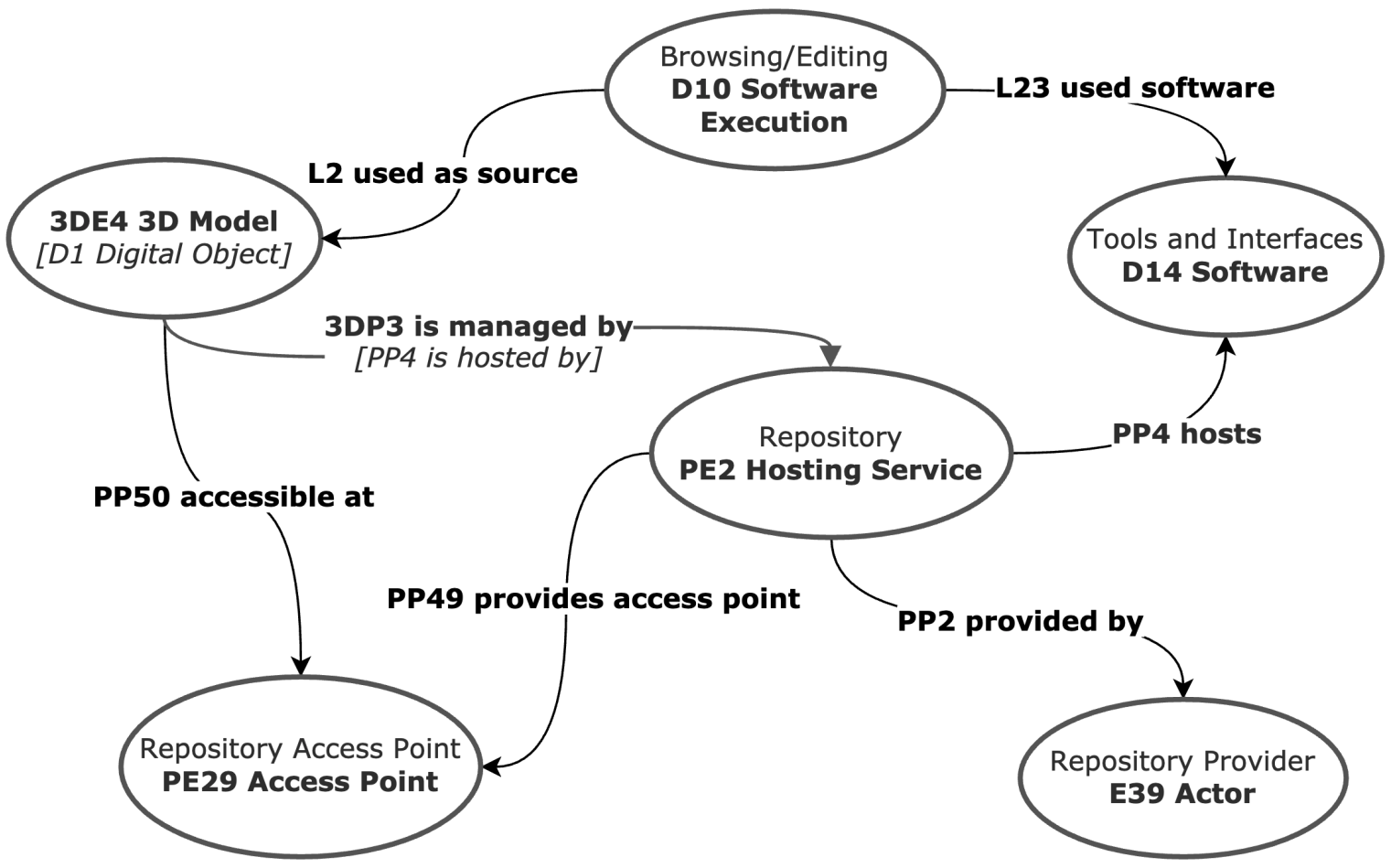

Fig. 7. Entities and properties for modelling the pubblication process.

## 4 METADATA AND ADDITIONAL INFORMATION

In the previous sections we have illustrated the various stages of the process on a conceptual level. However, it is necessary to consider the mechanisms and tools that researchers will use to efficiently generate, out of our application profile, information in a formal way, and encode it in ontological languages such as RDF or OWL.

Defining the process of metadata creation is of paramount importance and, among other things, it requires specific interfaces for the manual inclusion of detailed information and the establishment of procedures to automatically generate metadata from data embedded within the 3D models, produced by software and available in the related documentation.

For a complete and standardised metadata stream, it is also essential to define thesauri and other controlled vocabularies for assigning names and types to all the entities involved in the digitisation process. Besides the very popular general purpose services available today, such as DBpedia and Geonames, the Cultural Heritage sector is already equipped with a complete and articulated set of vocabularies designed for the description and identification of objects, monuments, places and temporal entities reputed of particular significance for the domain. The Getty Art & Architecture Thesaurus (AAT) [38], for example, besides providing concepts for each entity regarding its domain, is a constantly updated tool that has considerably extended its range of action to other disciplines as well. The PeriodO gazetteer of periods [44], on the other hand, is a fundamental resource for the definition of artistic and cultural periods and offers a smart environment based on the Linked Open Data paradigm and completely configurable by the user. The possibility of extending these terminological tools also to the 3D world by creating appropriate conceptual extensions is certainly of particular importance since thesauri constitute a privileged vehicle for data sharing and integration and their use is always strongly recommended.

### 4.1 Generating metadata for the 3D models

Most of the time, a consistent part of data relating to the digitisation process is not available within the 3D models, the software or elsewhere, but needs to be manually entered. Thus, for information concerning e.g. the time and place where the documentation and digitization processes were carried out, the people, institutions and projects involved and so on, it is essential to have simple and efficient interfaces equipped with advanced facilities that simplify data entry by minimizing user interaction. These interfaces become particularly important at ingestion time and need to be properly set up in order to establish and maintain a close relationship with the structure of the repositories where the digital models are physically stored.

A very similar problem has already been addressed for the cataloguing and storage of scientific data, for which a flexible and user-friendly interface, called THESPIAN-MASK [54], has been defined. THESPIAN-MASK is compatible with CIDOC CRM, provides all the necessary facilities to make the data entry process fast and efficient (e.g., autocomplete, reuse of previously inserted information, date pickers and so on) and is equipped with advanced features for LOD acquisition, able to query remote thesauri and gazetteer (such as Getty AAT and PeriodO) for the rapid inclusion of conceptual and spatial information (see Figure 8).

### 4.2 Deriving metadata from the 3D models

As we already mentioned, through a series of particular and sophisticated techniques it is possible to derive valuable information about 3D models, their content and their structure from the models themselves and the software used for their creation and editing. In a previous work [36] we have already examined the possibility of equipping 3D models with metadata for a better documentation of excavation sites and designed a plugin to be installed in software such as Blender, providing functions for interacting with 3D models and embedding semantic information directly inside its digital structure. Additionally, for the 3D models of sites, buildings and monuments created using BIM and HBIM, for example, it is possible to extract a huge amount of information from the IFC exchange format [15], a standard used by these systems to encode elements, features and relationships, and transform it into precious metadata. IFC mainly uses the ISO standard EXPRESS, a data specification language able to describe every aspect of the buildings represented in 3D, including structural and functional elements and the mutual relationships between their various components. Although it is not totally Web-compliant, the EXPRESS syntax is convertible, through appropriate procedures, to standard formats like RDF, OWL and languages of the semantic web. The possibility of extracting semantic information from IFC and building ontological structures starting from them has already been investigated and verified several times in recent years [22, 57, 58].

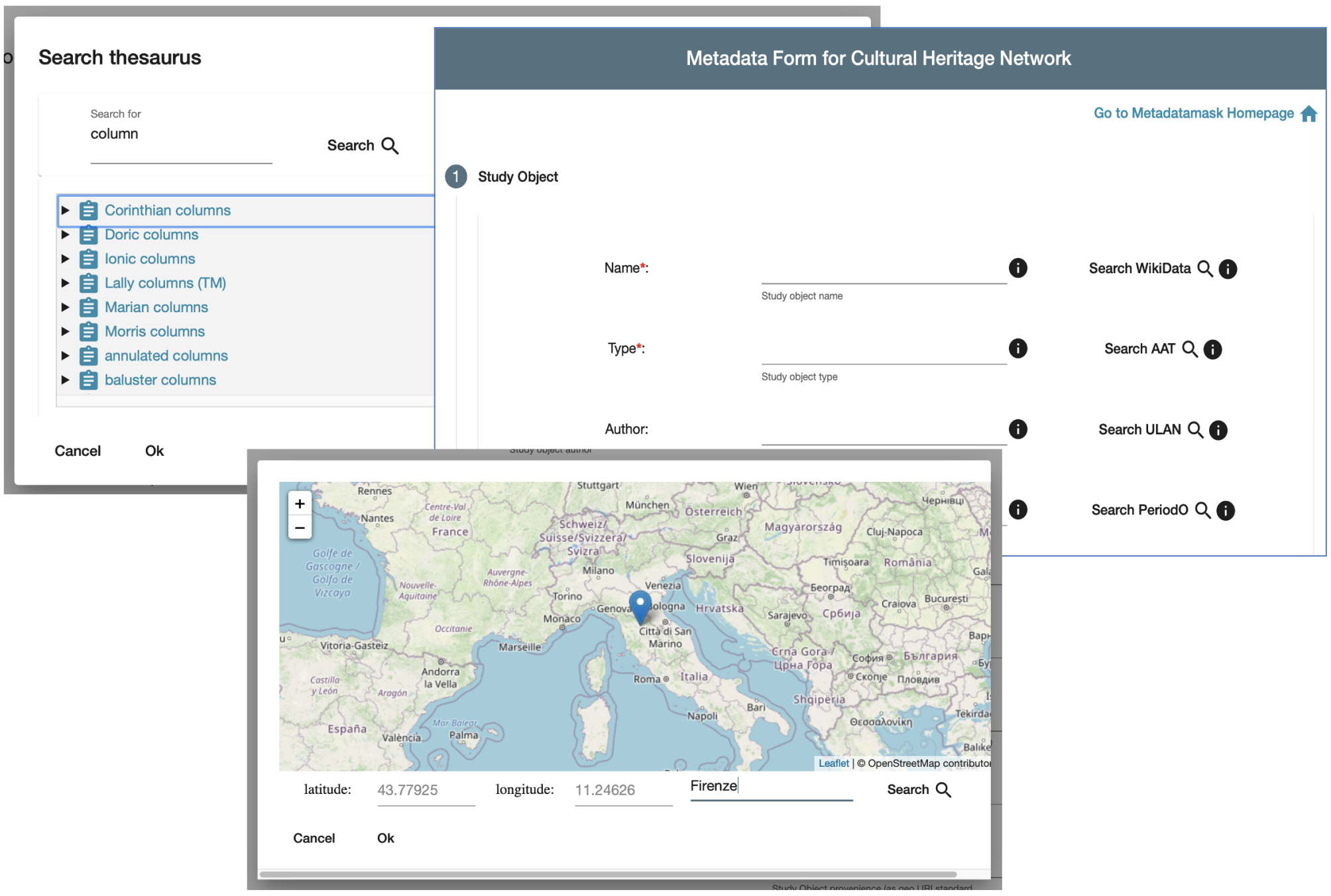

Fig. 8. THESPIAN-MASK interfaces to encode places, types and objects using gazetteers and thesauri in LOD format.

By employing these procedures it is possible to identify characteristic or relevant elements for the study of cultural sites and monuments and define new classes and properties out of them in order to construct an ontological extension capable of describing the relationships that these elements possess within the 3D models. Afterwards, it would be possible to compare and link these elements with those of the real world, for example with that of the archaeological buildings documented through CRMba. This possibility has also been explored by several scholars [18, 41, 48]. The results are encouraging but require further investigation, especially regarding the conceptual modelling and the integration of spatial, temporal, historical and artistic data to produce complete and effective tools.

## 4.3 Extracting metadata from textual documentation

3D models are often accompanied by a large amount of textual documentation containing information of great importance, but often ignored and almost never reported in metadata, usually describing models, their content and their creation. In addition to bibliographic information, of paramount importance in this regard is obviously the documentation gathered during the preparation phases, including archival documentation, reports of surveys and inspections, cost evaluations, tools assessment and so on. Additionally, there can be constellations of diaries and annotations drawn up during the different phases of the process, often using a natural language, which, if properly analysed and formalized, can prove to be excellent sources of knowledge.

In recent years, our team has developed advanced text analysis tools aimed at extracting this kind of knowledge from free texts and to define metadata in semantic format out of it. In particular TEXTCROWD [37, 53], created within the European Open Science Cloud initiative (EOSC), is an advanced cloud-based NLP tool powered by Supervised Machine Learning algorithms (Conditional Random Fields and Neural Networks) capable of identifying relevant entities such as objects, sites, monuments, events, periods and actors within excavation reports and other archaeological documentation, and to encode them in CIDOC CRM format. TEXTCROWD is also able to browse big online knowledge repositories and to train itself on demand to become quickly usable in different domains and with different languages.

Another newly developed cloud-based tool is THESPIAN-NER [54] that also uses deep learning techniques (namely a Convolutional Neural Network) for the natural language analysis of reports, annotations and similar texts produced by scientists and other scholars during laboratory analyses. The tool is designed for analysing texts to identify elements of the Heritage Science world, a branch of science aimed to improve the knowledge of Cultural Heritage entities. THESPIAN-NER allows the automatic classification of documents based on defined criteria and on controlled vocabularies, and to generate semantic information for the enrichment of existing knowledge graphs. The tool is in fact easy to integrate within existing metadata repositories and is able to interact with them both for the insertion of new information extracted from texts, and for their use as criteria to query and retrieve related existing data.

These tools can be employed also in our scenario to enable document classification and text analysis. The extracted information can be used to populate the repositories where 3D models are stored with annotated documentation, and to add descriptive metadata to their knowledge bases. The latter operation is of particular interest because it would foster the retrieval and reuse of important information concerning the digitization process often contained in textual documents such as preparatory reports, definition of user requirements and scenarios, but also observations, assessments and procedural indications reported by scholars and technicians along the whole 3D models creation process.

## 5 CONCLUSIONS AND FUTURE WORK

The research activity described in this paper has clearly shown how the use of ontologies and standards such as those proposed by the CIDOC CRM ecosystem can be specialized and combined to build tailor-made models to describe complex scenarios. CIDOC CRM, although it may seem complex in appearance, is instead a flexible system capable of identifying even the slightest conceptual nuances of various research domains and to profitably relate their various elements in logical and coherent ways. There are naturally many open issues that need to be addressed and additional work remains to be done to arrive at the formulation of an organic and complete model for 3D.

CRMdig, for instance, is now a few years old and, although it is still effective to describe 3D object scanning and model creation as in 2013, often appears rigid and not always adequate to render the complexity of modern digital systems and platforms. Despite the integration proposal provided by CRMpe and the many experiments to extend its classes and properties, attempted many times even in recent years (see for instance [26]), CRMdig would need a profound revision to make it a more flexible and effective tool.

The other CIDOC CRM extensions also present some critical issues especially in the way they are aligned, and are not always able to grasp the profound essence of the elements that revolve around the world of 3D.

Regarding the procedures for extracting metadata from 3D models, as discussed in Chapter 4, some preliminary experiments carried out by our team have verified the possibility of building a CIDOC CRM-enabled ontological extension for BIM/HBIM capable of automatically generating instances for its classes starting from IFC entities. By means of mapping and conversion scripts, it would be quite easy to define new classes representing 3D environments (such as *3DEx Site*, *3DEx Space*, *3DEx Bulding* and so on) and create instances for them derived from IFC elements such as *ifcSite*, *ifcSpace*, *ifcBuilding*, *ifcWall*. Using relationships such as *P67 refers to* or *P138 represents* it would then be possible to link the 3D entities to instances of CIDOC CRM and CRMba classes like *E27 Site*, *B2 Morphological Building Section*, *B1 Built Work*, *B4 Empty Morphological Building Section*, in order to establish precise semantic correspondences between elements of the digital and the real world. This would have a tremendous impact on the deep integration of 3D models within existing Cultural Heritage digital archives like the one developed by ARIADNEplus and the one that is going to be implemented by 4CH [17]. We will further explore the possibility to develop such an extension as future work.

As for the repositories, in addition to the existing ones, mainly focused on visualisation and sharing, it is of vital importance to design new platforms able to store digital data and provenance information in structured ways, since the quality of metadata plays a fundamental role in the establishment of integration and interoperability at a higher level. We have already collaborated in different cloud-based initiatives, such as EOSC and ARIADNEplus, and worked on the development of specific services for access and fruition of archaeological and scientific data. These systems, typically provide large space and calculation resources to be used for storing and processing 3D models. They also offer collaborative services and advanced interfaces, usually arranged in controlled environments such as virtual laboratories and virtual research environments (VRE), in which researchers can deploy and use dedicated tools to visualise and handle the available data. Our future research on this front will also focus on the design of new repositories and platforms capable of exploiting the FAIR principles to their full potential.

Another interesting topic to be faced concerns the fact that the CIDOC CRM system usually serve for the purpose of representing, in a formal way, activities that have already happened sometimes in the past. But planned activities, like the one typical of the preparation phases of 3D digitisation, are conceptually different from actual activities since they describe ways in which things are intended to happen, as opposed to ways in which things were actually carried out. For example, the *D2 Digitization Process* and *D28 Digital Documentation Process* classes of CRMdig are designed to document actions already performed and are not really suitable for the description of future activities.

Therefore, documenting a 3D digitisation plan would be of great benefit for the research, since it would allow establishing a formal description of all the necessary steps to be performed in the future by scholars, including details about the sequence of activities required, and all the people, objects, instrumentation and methodologies involved. Additionally, a semantic description of the digitisation process can be used to generate formal protocols for future works and also serve as a quality control framework to guarantee the reliability of the generated 3D data. Significant advances have already been made in other disciplines to model research processes [45, 46], but no suitable models still exist for the 3D domain.

A recent extension of the CIDOC CRM for social phenomena, still under definition and called CRMsoc [31], already offers the prototype of an *Activity Plan* class, subclass of *E29 Design or Procedure*, intended to model "plans foreseeing specific predefined activities or kinds of activities taking place and consisting of descriptions of specific constraints, patterns or types of activities that could be realized". CRMsoc also offers three properties for describing the events which the plan is intended for (*planned for*), the activities required for the plan to take place (*requires event of type*) and the documents which holds the assessment of the activity plan after it has been executed (*is assessed by*). These prototypical entities could be used as a base to build a mini ontology or extension for the documentation of all aspects of a 3D digitisation plan.

Since planning consists of a series of successive steps and sub steps and in the definition of the exact sequence in which these steps follow each other, it is therefore necessary to define specific sub classes and sub properties for modelling them and establishing coherent relationships for connecting each stage to the various planned activities it will consists of once it will be actually performed. This will allow modelling a complete workflow by decomposing the process plan into a series of separate, but linked, procedures. As part of our future activities, we will focus on defining the classes and properties of this new model and test it on real cases.

## 6 ACKNOWLEDGMENTS

The present research has been partially supported by 4CH, a project funded by the European Commission under the Horizon 2020 Program, Grant Agreement n.101004468.

## APPENDIX

Below we provide a general overview of the new classes and properties defined for our CRM3D application profile.

### *Classes*

**3DE1 Digital Model Creation**
Subclass of: E7 Activity (CIDOC CRM)
This class specializes the *E7 Activity* class of the CIDOC CRM and is intended to describe in general terms the complex of actions necessary for the creation of a digital model, such as those consisting in the survey of objects and places, research and analysis of sources, those of creating digital content and any other activity related to digitization. The activities of which a *3DE1* can be composed are identified through the property *P9 consists of* of the CIDOC CRM.

**3DE2 Digitisation Campaign**
Subclass of: E7 Activity (CIDOC CRM)
This is a subclass of *E7 Activity* used to describe campaigns, i.e., coordinated sets of activities planned by a project or a team to achieve specific goals. Typically, a campaign can last along a usually long time span (days, weeks or months), requires the work of several people and can be divided into working sessions. To relate a campaign with the various sessions of which it is composed, the *3DP2 has session* property is used.

**3DE3 Digitisation Session**
Subclass of: E7 Activity (CIDOC CRM)
This subclass of *E7 Activity* is used to model the sessions, i.e., those work actions, usually carried out within a short time frame (typically of hours or a day), within which the planning, analysis, evaluation, digitization and processing activities are carried out. The relationship between the set of activities for the creation of a digital model (*3DE1*) and the session within which they take place is expressed through the property *3DP1 was made during session.*

**3DE4 3D Model**
Subclass of: D1 Digital Object (CRMdig)
This class specializes the *D1 Digital Object* class of CRMdig to render in detail the concept of 3D model intended as a peculiar digital object having its definite identity and resulting from operations such as digitization, acquisition, processing and other actions typical of the three-dimensional modelling world. The peculiar features of a 3D model (e.g., its type, format, resolution, size, etc.) and its relationships with the series of activities carried out for its creation and manipulation are modelled through the properties inherited from its superclass (*D1*) and through the other classes and properties of CRMdig.

***Properties***

**3DP1 was made during**
Subproperty of: P10 falls within (CIDOC CRM)
Domain: 3DE1 Digital Model Creation
Range: 3DE3 Digitisation Session
This property links instances of model building activities (*3DE1*) with instances of the working sessions (*3DE3*) within which they occur.

**3DP2 has session**
Subproperty of: P9 consists of (CIDOC CRM)
Domain: 3DE2 Digitisation Campaign
Range: 3DE3 Digitisation Session
This property is used to model the relation between instances of *3DE2 Digitisation Campaign* and the working sessions (*3DE3*) taking place within their framework.

**3DP3 is managed by**
Subproperty of: PP4 is hosted by (CRMpe)
Domain: 3DE4 3D Model
Range: PE2 Hosting Service (CRMpe)
This property links instances of *3DE4 3D Model* with instances of *PE2 Hosting Service* (CRMpe) to describe the repository on which a model is stored and managed through specific services.

All web references retrieved on 30 April 2021.